\documentclass{emulateapj}
\usepackage{apjfonts}

\begin{document}

\title{A Long-Lived Accretion Disk Around a Lithium-Depleted Binary
  T Tauri Star}

\author{Russel J. White and Lynne A. Hillenbrand}

\affil{Department of Astronomy, California Institute of Technology, MS
  105-24, Pasadena, CA 91125}

\begin{abstract}

We present a high dispersion optical spectrum of St 34 and identify the
system as a spectroscopic binary with components of similar luminosity
and temperature (both M3$\pm$0.5).  Based on kinematics, signatures of
accretion, and location on an H-R diagram, we conclude that St 34 is a
classical T Tauri star belonging to the Taurus-Auriga T Association.
Surprisingly, however, neither component of the binary shows Li\,I 6708
\AA\, absorption, the most universally accepted criterion for establishing
stellar youth.  In this uniquely known instance, the accretion disk
appears to have survived longer than the lithium depletion timescale.  We
speculate that the long-lived accretion disk is a consequence of the
sub-AU separation companion tidally inhibiting, though not preventing,
circumstellar accretion.  Comparisons with pre-main sequence evolutionary
models imply, for each component of St 34, a mass of $0.37\pm0.08$
M$_\odot$ and an isochronal age of $8\pm3$ Myr, which is much younger than
the predicted lithium depletion timescale of $\sim 25$ Myr.  Although a
distance 38\% closer than that of Taurus-Auriga or a hotter temperature
scale could reconcile this discrepancy at 21-25 Myr, similar discrepancies
in other systems and the implications of an extremely old accreting
Taurus-Auriga member suggest instead a possible problem with evolutionary
models.  Regardless, the older age implied by St 34's depleted lithium
abundance is the first compelling evidence for a substantial age spread in
this region.  Additionally, since St 34's coeval co-members with early M 
spectral types would likewise fail the lithium test for youth, current
membership lists may be incomplete.

\end{abstract}


\newpage

\section{Introduction}

T Tauri stars are a class of young ($\lesssim$ 1-10 Myr) low mass ($\sim
0.1 - 2$ M$_\odot$) stars.  Those which show signatures of accretion from a
circumstellar disk, such as optical veiling and/or strong emission lines
(e.g. H$\alpha$) are called classical T Tauri stars while
those without accretion signatures are called weak-lined T Tauri stars.
Observational studies of these stars have provided the foundation upon
which current theories of star and planet formation are based.

Initial efforts to find T Tauri stars primarily relied on objective-prism
imaging surveys of dark clouds in search of strong emission line stars
\citep[e.g.][]{joy49}.  This technique, while relatively effective, biased
the discovered populations by mostly identifying classical T Tauri stars.
Subsequent surveys that focused on proper motion \citep[e.g.][]{jh79,
  hartmann91}, infrared excesses \citep[e.g.][]{kenyon90},
coronal/choromospheric indicators of youth such as x-ray emission
\citep[e.g.][]{walter88, wichmann96} and Ca\,II H\&K emission
\citep[e.g.][]{herbig86}, or location on an H-R diagram
\citep[e.g.][]{briceno98}, helped to establish a more complete and less 
biased census of star forming regions (most notably, Taurus-Auriga).
However, since some older binary star systems
(e.g. RS CVn type stars) and post-main sequence stars (e.g. AGB stars) also
exhibit many of these same properties, confirmation of T Tauri status
(i.e. extreme youth) has usually necessitated measurement of the
surface abundance of $^7$Li.  During the pre-main sequence (PMS)
contraction of a young star, $^7$Li is destroyed via p,$\alpha$ reactions
in the stellar interior when the central temperature rises above 
$\sim 3\times10^6$ K \citep{bodenheimer65}.  Because of rapid mixing in
fully convective low mass T Tauri stars, lithium is
completely depleted in a small fraction of the contraction timescale
\citep{dm94, bildsten97, bcah98, burke04}.  The presence of the strong,
easily observable Li\,I 6708 \AA\, absorption feature therefore implies
that the star must be very young.  The depletion timescale is a strong
function of mass, however, being quickest ($\sim 20$ Myr) for stars of mass
$\sim 0.6$ M$_\odot$.  The onset of a radiative core, which inhibits
efficient mixing, prior to full lithium depletion in higher mass stars
delays their depletion timescale.  The cooler central temperatures of lower
mass stars likewise delays their depletion timescale; lithium-burning
temperatures are never reaches for substellar objects with M $\lesssim
0.06$ M$_\odot$.  Thus at the highest and lowest T Tauri star masses, other 
diagnostics are needed to confirm a star's extreme youth.

Here we present high-dispersion spectroscopic observations of St 34 (HBC
425; RA: 04 54 23.7, DEC: +17 09 54, J2000; V$=14.4$ mag),
discovered as a strong 
H$\alpha$ emission line star in the objective prism survey of \citet{s86}
and later shown to be an early- to mid-M star \citep{dk88}.  Because of its 
strong H$\alpha$ emission and location, it has been assumed to be
a member of the Taurus-Auriga T Association \citep{kh95}.  Our new
measurements demonstrate that St 34 is a spectroscopic binary, a classical
T Tauri star, and strengthen the case for its association with
Taurus-Auriga.  Unlike all other classical T Tauri stars and known members
of Taurus-Auriga, however, St 34 has depleted its lithium.  These results
are used to assess the validity of evolutionary model predictions, the
possibility of a lithium-depleted population in Taurus-Auriga, and the
influence of sub-AU separation companions on circumstellar disk lifetimes.

\section{Spectroscopic Observations and Inferred Properties}

The W. M. Keck I 10-m telescope and High-Resolution Echelle Spectrometer
\citep{vogt94} were used on 2003 Feb 17 to obtain a high dispersion (R
$\approx$ 34,000) optical spectrum (6330-8750 \AA) of St 34.  The
observational setup, spectral calibration and extraction are as described
in \citet{wh04}.  Portions of the resulting spectrum of St 34 are shown in
Figure 1.  Its most distinguishing characteristics are strong, broad
H$\alpha$ emission, double-lined photospheric features (implying binarity),
and no Li\,I 6708 \AA\, absorption.  A Li\,I equivalent width upper limit
of 0.06 \AA\, for each spectroscopic component is determined by the size of
features in the psuedo-continuum.

\begin{figure}
\epsscale{1.2}
\plotone{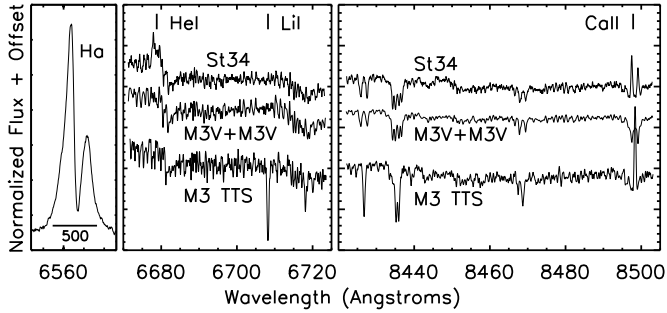}
\caption{Portions of the Keck/HIRES spectrum of St 34 showing the strong,
  broad H$\alpha$ emission profile (\textit{left panel}) and 2
  temperature sensitive regions (\textit{right panels}); the wavelength
  scales are the same in all panels.  The bar under the the H$\alpha$
  profile indicates a velocity width of 500 km/s.  A synthetic
  spectroscopic binary, composed of 2 M3 dwarf stars, and the M3
  weak-lined T Tauri star TWA 8a are also shown for comparison.  St 34
  shows no lithium in its spectrum.}
\end{figure}

The stellar properties of St 34 are determined from analysis of each
component's spectral features\footnote{High spatial
resolution imaging has identified a star at 1\farcs2
distance from St 34, possibly making the system a hierarchical triple
\citep{white04}.  This candidate companion is much fainter than the primary
pair at 2.2 $\mu$m ($\Delta K = 2.5$ mag), and thus unlikely to
contaminate the optical spectrum.}.  The procedure is
described in \citet{wh04}, but is tailored here for the analysis of a
spectroscopic binary.  Visual inspection of St 34's spectrum reveals that
the two components are of similar brightness, well separated in velocity,
and slowly rotating.  This allows a more independent analysis of each
component.  Radial velocities and $v$sin$i$ values are measured by fitting
the 2 peaks of the cross-correlation function determined using
non-rotating early- and mid-M dwarf templates.  Each component's spectral
type, the system flux ratio, and the continuum excess (or veiling, defined
as $r = F_{excess}/(F_{prim}+F_{sec})$) are determined simultaneously by
comparisons with synthetic spectroscopic binaries, generated by combining
dwarf standards at the appropriate radial and rotational velocities.  For
the synthetic spectra, the spectral type of each component is allowed to
vary from M0 to M5, the flux ratio from 0.5 to 2.0, and the continuum
excess from 0.0 to 2.0.  The best fit is determined by minimizing rms
differences between St 34 and a synthetic binary spectrum
over several temperature sensitive regions \citep[see][for details]{wh04}.
The components have the same spectral type (both M3$\pm$0.5) and are of
similar brightness.  The primary is assumed to be the star that is slightly
(10$\pm$4\%) brighter at 6500 \AA . Although the composite St 34 system
appears to have a small amount of continuum excess at 6500 \AA\, ($r_{6500}
= 0.13\pm0.05$), it is not possible to tell if this excess is associated
with only one or both of the components.  No continuum excess is detected
at longer wavelengths ($r_{8400} < 0.09$).  Weak, narrow He\,I 6678 \AA\,
and Ca\,II 8498, 8662 \AA\, emission is observed (Figure 1), but no
forbidden line emission (e.g. EW [SII]\,6716, 6731 \AA\, $<$ 0.05 \AA).
Table 1 summarizes the spectroscopic properties of St 34.

\section{Evidence for Adolescence}

The H$\alpha$ emission line profile of St 34 is both strong (equivalent
width = -51.6 \AA) and broad (full-width at 10\% of the peak = 512 km/s).
The breadth of this feature is not a consequence of binarity; the two peaks
in the profile are separated in velocity by $\sim 180$ km/s, which is much
greater than the velocity separation of the components (58.4 km/s).  
Although some late-type main sequence stars also display H$\alpha$
emission, caused by 
flares and other chromospheric activity, this emission is less intense and 
less broad than the profile of St 34 or any accreting young star
\citep[e.g.][]{wb03}.  Chromospherically active main-sequence stars also
tend to be rapidly rotating and X-ray bright, in contrast to the small
$v$sin$i$ values and X-ray non-detection of St 34 \citep[via
  ROSAT;][]{konig01}.  Moreover, the observed strength of the H$\alpha$
emission is similar to that seen in previous low spectral resolution
observations of St 34 \citep[EW$=-78$ \AA;][]{dk88,kh95}, 
suggesting the current epoch was not a transient flare-like event.
The H$\alpha$ emission line profile of St 34, in light of its stellar
properties, is most consistent with originating from a high velocity
accretion flow \citep[e.g.][]{muzerolle00}.  The low level continuum excess
at 6500 \AA\, supports the interpretation that one or both components of
the binary are accreting, though the accretion rate is low
\citep[$2.5\times10^{-10}$ M$_\odot$/yr, following a prescription similar
  to][]{wh04}.  Based on this evidence for accretion, we conclude that St
34 is a classical T Tauri star.

\begin{deluxetable}{lccc}
\tablewidth{0pt}
\tablehead{ \colhead{}
& \colhead{system} 
& \colhead{primary} 
& \colhead{secondary} }
\startdata
Obs. Julian Date	& 2452688.85	& \nodata	& \nodata \\
EW[H$\alpha$] (\AA)	& -51.6		& \nodata	& \nodata \\
H$\alpha$ 10\%-width (km/s)& 512	& \nodata	& \nodata\\
EW[Li\,I]	(\AA)	& \nodata	& $< 0.06$	& $< 0.06$ \\
$v$sin$i$ (km/s)	& \nodata	& $< 7.2$	& $< 7.0$	\\
Radial Velocity	(km/s)	& $17.9\pm0.6$	& $47.1\pm0.4$  & $-11.3\pm0.4$ \\
Spectral Type		& \nodata	& M3$\pm$0.5	& M3$\pm$0.5	\\
$[F_{prim}/F_{sec}]_{6500}$& $1.10\pm0.04$	& \nodata	& \nodata \\
$r_{6500}$		& $0.13\pm0.04$ & \nodata	& \nodata \\
$[F_{prim}/F_{sec}]_{8400}$& $0.99\pm0.07$	& \nodata	& \nodata \\
$r_{8400}$		& $< 0.09$	& \nodata	& \nodata \\
Mass (M$_\odot$)	&\nodata	& 0.37$\pm$0.08 & 0.37$\pm$0.08 \\
Isochronal Age (Myr)	&\nodata	& 8$\pm$3	& 8$\pm$3 \\
Lithium Depletion Age (Myr) &\nodata	& $>25$		& $>25$
\enddata 
\tablecomments{The systemic radial velocity is the average of the
components.  Its uncertainty does not reflect the uncertainties in
their masses.} \end{deluxetable}

The reservoir of accreting material has not yet been detected, however.
St 34 has no measurable near-infrared ($K_s$) excess, based on comparing
the observed $K_s$ magnitude to that predicted from the spectral type,
extinction (determined below), and $J$ magnitude.  Although \citet{wj92}
claim St 34 was detected by IRAS at $12 \mu$m, $25 \mu$m, and $60 \mu$m,
inspection of the IRAS Sky Survey Atlas, even after the more up-to-date
HiRes processing, reveals no point- or extended-source within several 
arcminutes of St 34.  We suggest the \citet{wj92} identification was
spurious.  St 34 was also not detected in the 1.3-mm survey by
\citet[][$F_{1.3mm} < 15$ mJy]{ob95}, which is sensitive to cool outer disk
material.  We emphasize, however, that the lack of detected excess emission
is consistent with St 34's low level accretion and cool stellar
temperature; several low mass accreting stars in Taurus-Auriga have not yet
been detected at far-infrared and millimeter wavelengths either
\citep{kh95}.

Kinematic information supports the assertion that St 34 is a member of the 
Taurus-Auriga T Association.  Assuming that the spectroscopic binary
components have the same mass, as suggested by their similar spectral
type and brightness, the systemic radial velocity is $17.9\pm0.6$ km/s, 
which is identical to the mean of Taurus members \citep[17.8
  km/s;][]{hartmann86}.  St 34 has a proper motion of $\mu_\alpha
=+1.8\pm3.5$ mas/yr and $\mu_\delta = -12.6\pm3.5$ mas/yr, which, in
combination with its radial velocity and 
an assumed distance of 145$\pm$10 pc, corresponds to space motion of
$U=-15.6\pm1.5, V=-8.4\pm2.2, W=-8.8\pm2.3$ km/s (E. Mamajek,
priv. comm.).  This motion is statistically most consistent with that of
lithium-rich Taurus members \citep[e.g.][]{jh79} as opposed to nearby
moving groups \citep[e.g. $\beta$ Pictoris;][]{zs04}.  Thus, based on
evidence for extreme youth, spatial proximity (within $1.5^\circ$ of
L1558), and space motion, we conclude St 34 is a member of the
Taurus-Auriga T Association.

Mass and age estimates for St 34 are determined by comparing the
temperature and luminosity with the \citet{bcah98} PMS evolutionary
models.  A temperature of 3415$^\circ$ K is assigned using the 
spectral type - temperature scale of \citet{luhman03}, which is slightly
hotter than a typical dwarf temperature scale, but yields coeval cluster
populations in combination with this evolutionary model.  A visual
extinction of 0.24 mag is determined by comparing the $J-H$ color to that
expected for an M3 star \citep{km94}, using a standard interstellar
extinction law \citep {rl85}.  The 2MASS magnitudes of St 34 are $J =
10.69\pm0.02$, $H = 10.08\pm0.02$, and $K_s = 9.79\pm0.02$; no reliable
optical colors are available.  Luminosity is computed by applying a
bolometric correction of $+1.75$ to the reddening corrected $J$ magnitude,
which is split assuming equal contribution from each component of the
binary.  A distance of 145$\pm$10 pc, corresponding to that of Taurus
\citep{bertout99}, yields log(L/L$_\odot$) = $-1.03\pm0.06$ for each
component.  In the top panel of Figure 2, St 34 is shown on an H-R diagram
along with the \citet{bcah98} evolutionary models.  For each component, the
implied age is $8\pm3$ Myr and mass is $0.37\pm0.08$ M$_\odot$.  St 34
appears to be somewhat older than most stars in Taurus, which have an
average age of 2-3 Myr \citep{wg01} and are all thought to be younger than
4 Myrs \citep{hartmann03}.

In the bottom panel of Figure 2, the lithium abundance of St 34 is
compared to the lithium depletion predictions of the same \citet{bcah98}
evolutionary models.  The Li 6708 \AA\, equivalent width of $\lesssim 0.06$
\AA\, corresponds to a lithium abundance of log$n$(Li) $\lesssim 0.3$ dex,
following the curves of growth shown in \citet[][Figure 2]{song02}.
This amount of depletion implies an age $\gtrsim 25$ Myr.  Comparisons
with other evolutionary models yield lithium depletion ages that agree
to within 20\% \citep[see e.g.][]{burke04}.

\section{Discussion and Implications}

\subsection{A Possible Problem for Lithium Depletion Predictions}

Although the isochronal age inferred for St 34 is much younger than its
lithium depletion age, a distance of 90 pc instead of the assumed 145 pc
would increase the isochronal age to 25 Myr, consistent with the lithium
depletion age.  Similarly, if the assumed temperature is increased to
3600$^\circ$ K (which is 340$^\circ$ K hotter than a typical M3 dwarf 
temperature), the isochrone and lithium depletion ages agree at 21 Myr.
At an age $\gtrsim 20$ Myrs, however, St 34 would be by far the oldest
classical T Tauri star known; MP Mus in Scorpius-Centaurus \citep[age $\sim
13$ Myr;][]{mamajek02} is currently thought to be one of the oldest.  St 34
would also be older than typical cloud dispersal timescales 
\citep[$< 10-20$ Myr;][]{pg97}, which would call into question its
association with Taurus-Auriga as location and kinematics
suggest.  This large age for St 34 seems unlikely.  \citet{song02}
identified a similar discrepancy between isochronal and lithium depletion
ages in the case of HIP 112312 A.  The low lithium abundance of this M4
star implies an age $\gtrsim 35$ Myr, while its isochronal age, which is
based on a Hipparcos determined distance, is $6\pm3$ Myrs
(Figure 2).  As with St 34, a hotter temperature could 
reconcile these ages at 20-25 Myr.  Finally, we note that the lithium
depletion ages of young open clusters, including the Pleiades, $\alpha$
Persei, IC 2391, and NGC 2547 \citep{burke04}, are all systematically
larger than the isochronal ages fitted to both low mass unevolved members
and the upper main sequence near the nuclear turn-off \citep{stauffer01,
  jn01}.  Overall, the emerging observational evidence suggests a problem 
with lithium depletion ages, being systematically too old, though a problem
with the pre-main sequence temperature scale, being too cool, can not be
ruled out.

\begin{figure}
\epsscale{1.2}
\plotone{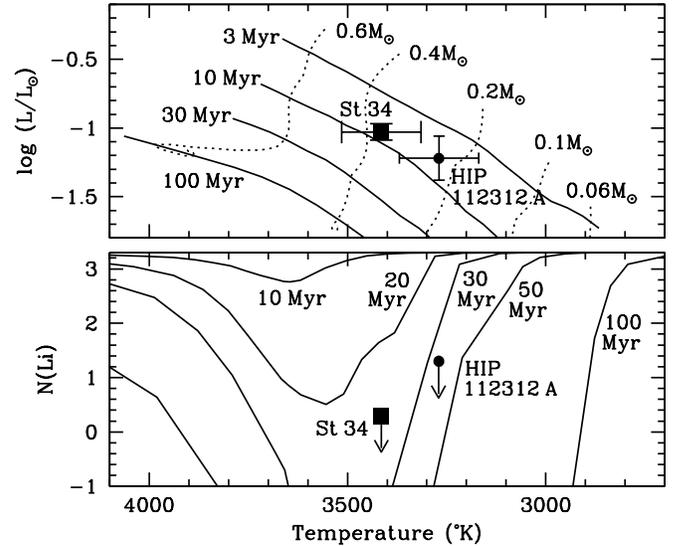}
\caption{In the top panel, St 34 (\textit{square}) and HIP 112312\,A
  (\textit{circle}) are shown on an H-R diagram along with the evolutionary
  models of \citet{bcah98}.  Since both components of St 34 have the same
  temperature and luminosity, only 1 point is visible.
  In the bottom panel St 34 and HIP 112312\,A are shown on a lithium
  abundance versus temperature diagram, along with the isochrones predicted
  by the same \citet{bcah98} models.  Although the location of St 34 and
  HIP 112312\,A on the H-R diagram suggests ages $\lesssim 10$ million
  years, the depleted lithium abundance of both stars suggests ages older
  than $\sim 25$ Myr and $\sim 35$ Myr, respectively.
}
\end{figure}

\subsection{Implications for the Taurus Population}

If St 34 is a member of the Taurus-Auriga T Association, as evidence
suggests, it presents the first strong case for a significant age spread in
this region.  St 34 must be older than currently forming stars by an amount
equal to the lithium depletion timescale.  Evolutionary models suggest this
could be as large as 21-25 Myr, but as noted above, these values appear
to be extreme.  The age spread would be more accurately constrained if
additional older, possibly lithium depleted members are discovered.  St 34
was easily identified because of its strong H$\alpha$ 
emission.  Lithium depleted weak-lined T Tauri stars, on the other hand,
even if identified and observed spectroscopically, would have thus far been
dismissed as non-members.  In order to estimate how many older weak-lined
systems there could be in Taurus, we use the ratio of weak-lined to
classical T Tauri stars in clusters of age $\sim 10$ Myr, which range from
$\sim 90$\% for the TW Hydrae Association \citep[age $\sim 8$ Myr;][]{zs04}
to $\sim 99$\% for the Sco-Cen subgroups \citep[age $\sim 13$
  Myr;][]{mamajek02}.  If surveys of these clusters are also biased by 
requiring the presence of lithium absorption for a member to be confirmed,
these ratios could be even higher.  This suggests that St 34 may have 10
and possibly many 10s of coeval co-members that are weak-lined T Tauri
stars.  Those of early-M spectral type will have likely depleted their
lithium (Figure 2).  Current membership lists of Taurus may therefore be
incomplete.

The Einstein and ROSAT surveys may have identified some of these lithium
poor stars.  \citet{walter88} and \citet{wichmann96} together identified 17
stars with spectral types M1 - M3.5 in Taurus, nearly half of which have
been subsequently dismissed as non-members because of depleted lithium.
Some of these lithium-poor stars nevertheless have radial velocities
consistent with Taurus \citep[e.g. RX J0446.8+2255;][]{wichmann00},
suggesting they could in fact be bona fide members.  Confirmation
of membership will require more accurate estimates of distance, proper
motion, and surface gravity.  We note that if this proposed older
population is identified, it has significant implications for the duration
of star formation in Taurus, which is generally believed to be less than
$4$ Myr (Hartmann 2003, but see Palla \& Stahler 2002), and for the initial
mass function of Taurus, which apparently peaks at a mass \citep[$\sim 0.8$
  M$_\odot$;][]{luhman03} slightly larger than the mass where lithium
depletion occurs first.

\subsection{Long-Lived Accretion Disks in Close Binary Systems}

With an age of $\gtrsim 8$ Myr, St 34 is one of a handful of old ($\gtrsim
10$ Myr) classical T Tauri stars.  We speculate than in many cases these
long-lived accretion disks are a consequence of a tidally inhibited
accretion flow caused by a sub-AU separation companion.  The properties of
St 34, for example, imply a binary separation of $\lesssim 0.78$ AU.  As
has been identified in some spectroscopic binary systems \citep[e.g. DQ
Tau;][]{basri97}, and predicted 
by numerical simulations \citep{as96}, the orbital dynamics of a close
binary does not preclude accretion from a circumbinary disk.  Nevertheless,
we suggest that it is less efficient.  In support of this, there is some 
evidence for a higher frequency of spectroscopic binaries among old
classical T Tauri stars than among younger T Tauri populations, for which
the binary fraction is only $7\pm3$\% \citep[for periods less than 100 
  days;]{mathieu94}.  Of the 4 classical T Tauri stars in the TW Hydrae
Association, 50\% are spectroscopic binaries \citep[TWA 5A, Hen
  3-600A;][]{muzerolle00, mohanty03}.  Of the remaining 2 accreting stars  
(TW Hya, TWA 14), only TW Hya has multiple high-dispersion measurements
sensitive to radial velocity variations, but its pole-on orientation
\citep{weinberger02} would significantly inhibit the detection of a close
companion in a coplanar orbit.  Both TWA 14 and TW Hya could be yet
unidentified sub-AU binary systems.  A more complete binary census of these
and other old accretors \citep[e.g. MP Mus;][]{mamajek02} is needed to
confirm this hypothesis.  One interesting implication is that planets would
have a longer time to form in the circumbinary disk of sub-AU separation
binary stars than around single stars.

\acknowledgements

We thank I. Baraffe, A. Ghez, and J. Stauffer for helpful discussions and
are grateful to E. Mamajek for generously providing valuable kinematic
information and insight.  We appreciate the data provided by the NASA/IPAC
Infrared Science Archive and the privilege to observe on the revered summit
of Mauna Kea.


\begin{thebibliography}{}

\bibitem[Artymowicz \& Stephen (1996)] {as96} Artymowicz, P. \& Lubow,
  S. H. 1996, \apj, 467, 77

\bibitem[Baraffe et al.(1998)] {bcah98} Baraffe, I., Chabrier, G.,
  Allard, F., \& Hauschildt, P. H. 1998, A\&A, 337, 403

\bibitem[Basri et al. (1997)] {basri97} Basri, G., Johns-Krull, C. M. \&
  Mathieu, R. D. 1997, \aj, 114, 781

\bibitem[Bertout et al. (1999)] {bertout99} Bertout, C., Robichon, N. \&
  Arenou, F. 1999, \aap, 352, 574

\bibitem[Bildsten et al. (1997)] {bildsten97} Bildsten, L., Brown, E. F.,
  Matzner, C. D. \& Ushomirsky, G. 1997, \apj, 482, 442

\bibitem[Bodenheimer (1965)] {bodenheimer65} Bodenheimer, P. 1965, \apj,
  142, 459

\bibitem[Brice\~no et al. (1998)] {briceno98} Brice\~no, C., Hartmann, L.,
  Stauffer, J. \& Mart\'\i n, E. 1998, \aj, 115, 2074

\bibitem[Burke et al. (2004)] {burke04} Burke, C. J., Pinsonnealult,
  M. H. \& Sills, A. 2004, \apj, 604, 272

\bibitem[D'Antona \& Mazzitelli (1994)] {dm94} D'Antona, F. \& Mazzitelli,
  I. 1994, \apjs, 90, 467


\bibitem[Downes \& Keyes (1988)] {dk88} Downes, R. A. \& Keyes, C. D. 1988,
  \aj, 96, 777

\bibitem[Gomez et al. (1993)] {gomez93} Gomez, M., Hartmann, L., Kenyon,
 S. J. \& Hewett, R. 1993, \aj, 105, 1927

\bibitem[Hartmann et al.(1986)] {hartmann86} Hartmann, L., Hewett, R.,
  Stahler, S., \& Mathieu, R. D., 1986, \apj, 309, 275

\bibitem[Hartmann et al.(1991)] {hartmann91} Hartmann, L., Stauffer, J. R.,
  Kenyon, S. J. \& Jones, B. F. 1991, \aj, 101, 1050

\bibitem[Hartmann (2003)] {hartmann03} Hartmann, L. 2003, \apj, 585, 398

\bibitem[Herbig et al. (1986)] {herbig86} Herbig, G. H., Vrba, F. J. \&
Rydgren, A. E. 1986, \aj, 91, 575

\bibitem[Jeffries \& Naylor (2001)] {jn01} Jeffries, R. D., \& Naylor,
  T. 2001, in ASP Conf. Ser. 243, From Darkness to Light: Origin and
  Evolution of Young Stellar Clusters, ed. T. Montmerle \& P. Andr\'e (San
  Francisco: ASP), 633 

\bibitem[Jones \& Herbig (1979)] {jh79} Jones, B. F. \& Herbig, G. H. 1979,
  \aj, 84, 1872

\bibitem[Joy (1949)] {joy49} Joy, A. H. 1949, \apj, 110, 424

\bibitem[Kenyon et al. (1990)] {kenyon90} Kenyon, S. J., Hartmann, L. W.,
 Strom, K. M. \& Strom, S. E. 1990, \aj, 99, 869

\bibitem[Kenyon \& Hartmann (1995)] {kh95}
  Kenyon, S. J. \& Hartmann, L. 1995 

\bibitem[Kirkpatrick \& McCarthy(1994)] {km94} Kirkpatrick, J. D. \&
  McCarthy, D. W. Jr. 1994, \aj, 107, 333

\bibitem[K\"onig et al. (2001)] {konig01} K\"onig, B., Neuh\"auser, R. \&
  Stelzer, B. 2001, \aap, 369, 971

\bibitem[Luhman et al. (2003)] {luhman03} Luhman, K. L., Brice\~no, C.,
  Stauffer, J. R., Hartmann, L., Barrado y Navascués, D. \& Caldwell,
  N. 2003, \apj, 590, 348

\bibitem[Mamajek (2002)] {mamajek02} Mamajek, E. E., Meyer, M. R. \&
  Liebert, J. 2002, \aj, 124, 1670

\bibitem[Mathieu (1994)] {mathieu94} Mathieu, R. D. 1994, \araa, 32, 465 

\bibitem[Mohanty et al. (2003)] {mohanty03} Mohanty, S., Jayawardhana,
 R. \& Barrado y Navascu\'es, D. 2003, \apj, 593, 109

\bibitem[Muzerolle et al. (2000)] {muzerolle00} Muzerolle, J., Calvet, N.,
  Briceño, C., Hartmann, L. \& Hillenbrand, L. 2000, \apj, 535, 47


\bibitem[Osterloh \& Beckwith(1995)] {ob95} Osterloh, M. \& Beckwith,
S. V. W. 1995, \apj, 439, 288

\bibitem[Palla \& Galli (1997)] {pg97} Palla, F. \& Galli, D 1997, \apj,
  476, 35

\bibitem[Palla \& Stahler (2002)] {ps97} Palla, F. \& Stahler, S. W. 2002,
  \apj, 581, 1194


\bibitem[Rieke \& Lebofsky(1985)] {rl85} Rieke, G. H. \& Lebofsky,
  M. J. 1985, \apj, 288, 618


\bibitem[Song et al. (2002)] {song02} Song, I., Bessell, M. S. \&
  Zuckerman, B. 2002, \apj, 581, 43

\bibitem[Stauffer et al. (2001)] {stauffer01} Stauffer, J. R., Jeffries,
  R. D., Mart\'\i n, E. L., \& Turndrup, D. M. 2001, in ASP Conf. Ser. 223,
  Cool Stars, Stellar Systems and the Sun, Ed. R. J. Garcia L\'opez,
  R. Rebolo, \& M. R. Zapatero Osorio (San Francisco: ASP), 399

\bibitem[Stephenson (1986)] {s86} Stephenson, C. B. 1986, \apj, 300, 779

\bibitem[Vogt et al. (1994)] {vogt94} Vogt, S. S., et al. 1994, \procspie,
  2198, 362

\bibitem[Walter et al. (1988)] {walter88}
Walter, F. M., Brown, A., Mathieu, R. D., Myers, P. C. \& Vrba,
F. J. 1988, \aj, 96, 297

\bibitem[Weaver \& Jones (1992)] {wj92} Weaver, W. B. \& Jones, G.
  1992, \apjs, 78, 239


\bibitem[Weinberger et al. (2002)] {weinberger02} Weinberger, A. J. et al.
 2002, \apj, 566, 409

\bibitem[White \& Basri (2003)] {wb03} White, R. J. \& Basri, G. 2003,
  \apj, 582, 1109

\bibitem[White \& Ghez (2001)] {wg01} White, R. J. \& Ghez, A. M. 2001, 

\bibitem[White \& Hillenbrand (2004)] {wh04} White, R. J. \& Hillenbrand,
  L. A. 2004, \apj, accepted

\bibitem[White et al. (2004)] {white04} White, R. J. et al. 2004, in prep.

\bibitem[Wichmann et al. (1996)] {wichmann96} Wichmann, R. et al. 1996,
  \aap, 312, 439

\bibitem[Wichmann et al. (2000)] {wichmann00} Wichmann, R. et al. 2000,
  \aap, 359, 181

\bibitem[Zacharias et al. (2004)] {zacharias04} Zacharias, N., Urban,
  S. E., Zacharias, M. I., Wycoff, G. L., Hall, D. M., Monet, D. G. \&
  Rafferty, T. J. 2004, \aj, 127, 3043

\bibitem[Zuckerman \& Song (2004)] {zs04} Zuckerman, B. \& Song, I. 2004, 
  \araa, 42, 685

\end{thebibliography}
\end{document}